\documentclass{iau-JDSS}
\usepackage{graphicx}

\title[JD8: Hot Interstellar Matter in Elliptical Galaxies] 
{Confronting feedback simulations with observations of hot gas in elliptical galaxies}

\author[Q. Daniel Wang]   
{Q. Daniel Wang$^1$}

\affiliation{$^1$Department of Astronomy, University of Massachusetts,\\
  Amherst, MA 01003, USA  email: {\tt   wqd@astro.umass.edu}}

\pubyear{2009}
\volume{xxx}  
\pagerange{}
\jname{Title of your IAU Symposium}
\editors{Dong-Woo Kim. Editor}
\begin{document}

\maketitle


Elliptical galaxies comprise primarily old stars, which collectively generate 
a long-lasting feedback via stellar mass-loss and Type Ia SNe. This 
feedback can be traced by X-ray-emitting hot gas in and around such galaxies,
in which little cool gas is typically present. However,
the X-ray-inferred mass, energy, and metal abundance of the hot gas
are often found to be far less than what are expected from 
the feedback, particularly in so-called low $L_X/L_B$ ellipticals.
This ``missing'' stellar feedback is presumably lost in galaxy-wide outflows,
which can play an essential role in galaxy evolution (e.g.,
explaining the observed color bi-modality of galaxies). 
We are developing a model that can be used to properly interpret
the X-ray data and to extract key information about the dynamics of 
the feedback and its interplay with galactic environment.

First, we have constructed a 1-D model of the stellar feedback in the 
context of galaxy formation and evolution. The
feedback is assumed to consist of two primary phases:
1) an initial burst during the bulge formation and 2) a subsequent
long-lasting mass and energy injection from stellar winds and Type Ia SNe
of low-mass stars. An outward blastwave is initiated by the
burst and is maintained and enhanced by the long-lasting stellar feedback.
This blastwave can heat the surrounding medium not only in the
galactic halo, but also in regions beyond the virial radius.  As a result, the
smooth accretion of hot gas can be completely stopped.
The long-lasting feedback can form a galactic bulge wind, which is
reverse-shocked at a large radius, and can later evolve into a subsonic
quasi-stable outflow as the energy injection decreases with time.
The two phases of the feedback thus re-enforce each-other's impact
on the gas dynamics. Present-day elliptical galaxies with significant amounts of
hot gas are most likely in subsonic outflow states. The exact 
properties of such an outflow depend on the galaxy formation history and
environment. This dependence and variance may explain the large dispersion 
in the $L_X/L_B$ ratios of elliptical galaxies. 

Second, to quantitatively compare the simulations with X-ray observations, we have conducted various 3-D hydrodynamical simulations with the adaptive mesh refinement code FLASH to investigate the physical properties of hot gas in and around elliptical galaxies.
We have developed an embedding scheme of individual supernova remnant seeds, which 
allows us to examine, for the first time, the effect of sporadic SNe on 
the density, temperature, and iron ejecta distribution of the hot gas 
as well as the resultant X-ray morphology and spectrum. We find that the SNe produce a wind/outflow with highly filamentary density structures and patchy ejecta. Compared with a 1-D spherical wind model, the non-uniformity of simulated gas density, temperature, and metallicity substantially alters the spectral shape and increases the diffuse X-ray luminosity. The differential emission measure as a function of temperature of the simulated gas exhibits a log-normal distribution, with a peak value much lower than that of the corresponding 1D model. The bulk of the X-ray emission comes from the relatively low temperature and low abundance gas shells associated with SN blastwaves. SN ejecta are not well mixed with the ambient medium and typically remain
very hot in the central region. Driven by the buoyancy, the iron-rich gas on average 
moves substantially faster than the medium and only gradually mixes with it on the way
out. As a result, apparent increasing temperature and metal abundance with off-center distance can arise in the region, mimicking what have been observed in elliptical galaxies. These results, at least partly, account for the apparent lack of evidence for iron enrichment in the soft X-ray-emitting gas in galactic bulges and intermediate-mass elliptical galaxies. More details will be presented in Tang \& Wang (2010).

\end{document}